\documentclass[12pt]{article}

\usepackage{graphicx}
\usepackage{epsfig}

\begin{document}
\renewcommand{\textfraction}{0}

\title{Layered Orthogonal Lattice Detector for Two Transmit Antenna Communications\footnote{This research
was supported by a Marie Curie International Fellowship within the
$6^{th}$ European Community Framework Programme}}
\author{\normalsize
Massimiliano Siti \\
\small Advanced System Technologies \\[-5pt] \small STMicroelectronics
\\[-5pt]
\small 20041 Agrate Brianza (Milan) - Italy \\[-5pt] \small
massimiliano.siti@st.com \and
\normalsize Michael P. Fitz \\
\small Dept. of Electrical Engineering \\ [-5pt]
\small University of California Los Angeles \\ [-5pt]
\small Los Angeles, CA 90095-1594\\ [-5pt]
\small fitz@ee.ucla.edu }
\date{}
\maketitle
\thispagestyle{empty}
\begin{abstract}
A novel detector for multiple-input multiple-output (MIMO)
communications is presented. The algorithm belongs to the class of
the {\it lattice detectors}, i.e. it finds a reduced complexity
solution to the problem of finding the closest vector to the
received observations. The algorithm achieves optimal
maximum-likelihood (ML) performance in case of two transmit
antennas, at the same time keeping a complexity much lower than
the exhaustive search-based ML detection technique. Also,
differently from the state-of-art lattice detector (namely {\it
sphere decoder}), the proposed algorithm is suitable for a highly
parallel hardware architecture and for a reliable bit soft-output
information generation, thus making it a promising option for
real-time high-data rate transmission.
\end{abstract}
\normalsize

\section{Introduction}
Wireless transmission through multiple antennas, also referred to
as MIMO (Multiple-Input Multiple-Output), currently enjoys great
popularity because of the demand of high data rate communication
from multimedia services.

In MIMO fading channels ML detection is desirable to achieve
high-performance, as this is the optimal detection technique in
presence of additive Gaussian noise. ML detection involves an
exhaustive search over all the possible sequences of digitally
modulated symbols, which grows exponentially as the number of
transmit antennas. Because of their reduced complexity,
sub-optimal linear detectors like Zero-Forcing (ZF) or Minimum
Mean Square Error (MMSE) \cite{Wolniansky} are widely employed in
wireless communications. Such schemes yield a low spatial
diversity order: for a MIMO system with $L_{t}$ transmit and
$L_{r}$ receive antennas this is equal to $L_{r}-L_{t}+1$, as
opposed to $L_{r}$ for ML \cite{Van Nee}. ZF and MMSE have also
been proposed in combination with Interference Cancellation (IC)
techniques \cite{Foschini}. However the performance of such
nonlinear detectors is better than linear detectors but does not
always give near-ML performance.

Lattice decoding algorithms, like sphere decoder (SD)
\cite{Viterbo}, have been proposed for systems whose input-output
relation can be represented as a real-domain linear model
\begin{equation}
\textbf{y}_r=\textbf{Bx}_r+\textbf{n}_r \label{sys_eq}
\end{equation}
where if $n=2L_{r}$ and $m=2L_{t}$, the channel output vector
$\textbf{y}\in R^{n}$, the input vector $\textbf{x}\in R^{m}$ is
carved from a discrete finite set of values and \textbf{B} is a
$n$ x $m$ real matrix representing the channel mapping of the
transmit {\it codebook} into a received {\it lattice} corrupted by
the Gaussian noise $\textbf{n}\in R^{n}$; $\textbf{B}$ is also
referred to as the {\it lattice generator} matrix. SD can attain
ML performances with significant reduced complexity. The lattice
formulation for MIMO wireless systems was described in
\cite{Damen} in case of QAM digitally modulated transmitted
symbols; in that case a system equation in the form (\ref{sys_eq})
can be derived.

Besides SD, to our knowledge the class of ML-approaching
algorithms is quite limited. Other examples include the reduced
search set presented in \cite{Rupp-Gritsch}, which does not yield
good performance below $10^{-4} BER$, or the approximate method
\cite{Sung-Lee}, which entails high complexity; also, no
performance results are reported for a constellation size larger
than QPSK.

The SD algorithm converges at the ML solution while searching a
much lower number of lattice points than the exhaustive search
required by a "brute-force" ML detector. However, it presents a
number of disadvantages; most important are:

\begin{enumerate}
  \item {It is an inherently \emph{serial} detector and thus is not
suitable for a parallel implementation.}
   \item {Parameter sensitivity.}  The number of lattice points to be
searched is variable and sensitive to many parameters like the
choice of the initial radius; the signal to noise ratio (SNR); the
(fading) channel conditions. This means it could be unsuitable for
applications requiring a real-time response in data
communications.
  \item {Bit soft output generation.} In \cite{ten_Brink} the idea of building
a "candidate list" of sequences to compute the bit log-likelihood
ratios (LLR) was discussed. Unfortunately the optimal size of such
a list is a function of the system parameters and can still be
very high (thousands of lattice points) for practical
applications.
\end{enumerate}

In this paper, we propose a novel layered orthogonal lattice
detector (LORD) for two transmit antenna MIMO systems, which
achieves ML performance in case of hard-output demodulation and
optimally computes bit LLRs when soft output information is
generated. Similarly to SD, LORD consists of three different
stages, namely a lattice formulation, different from the one
introduced in \cite{Damen} and typically used by SD; the
preprocessing of the channel matrix, which is basically an
efficient way to perform a QR decomposition; and finally the
lattice search, which finds an optimal solution to the
\emph{closest vector} problem \cite{Agrell}, given the
observations. The innovative concept, compared to SD, is that the
search of the lattice points can be made in a parallel fashion,
and fully deterministic. The number of lattice points to be
searched is well below the exhaustive search ML algorithm, and for
soft output generation is linear in the number of transmit
antennas.

The paper is organized as follows. In Section \ref{sys} we
introduce the system notation used throughout the paper and
describe the lattice representation for§ LORD. In Section
\ref{preproc} the preprocessing algorithm of the lattice matrix is
described. Section \ref{ML_demod} details the reduced complexity
ML demodulation technique. Its principles lead to the formulation
of the optimal \emph{max-log} bit LLR derivation, explained in
Section \ref{LLR}. Section \ref{perf} shows the performance
results obtained applying LORD to a BICM system and flat Rayleigh
fading channel. Finally, Section \ref{concl} concludes the paper.

\section{System Notation and Lattice Formulation} \label{sys}
The scenario considered in this document is a linear MIMO
communication system with $L_t=2$ transmit and $L_r$ receive
antennas and frequency nonselective fading channel. The
information symbol vector \textbf{x} $=(X_1 \, \, X_2)^{T}$, where
$X_j,\,j=1,\,2$ is a complex symbol belonging to a given
quadrature-amplitude modulation (QAM) or phase-shift keying (PSK)
constellation, is distributed among the two transmit antennas and
synchronously transmitted. The signal received at each antenna is
therefore a superposition of the two transmitted signals corrupted
by multiplicative fading and additive white Gaussian noise (AWGN).
The output of the matched filters to the pulse shape at each
receive antenna can be written in matrix notation as:

\begin{equation}
\textbf{y}=\sqrt{\frac{E_s}{2}} \, \textbf{Hx}+\textbf{n}
\label{mimomatr1}
\end{equation}
where $E_s$ is the energy per transmitted symbol (under the
hypothesis that the average constellation energy is
$E[|X_j|^2]=1$); the entries of the $L_r$ x 2 channel matrix
\textbf{H}, $H_{ji}$, represent the complex path gains from
transmit antenna $i$ to receive antenna $j$; \textbf{y} $=(Y_1 \,
\ldots Y_{L_r})^{T}$ and \textbf{n} $=(N_1 \, \ldots N_{L_r})^{T}$
are the $L_r$ x 1 complex received signal and AWGN sample vectors
respectively. The complex path gains are samples of zero mean
Gaussian random variables (RVs) with variance $\sigma^2_H =0.5$
per real dimension. Fading processes for different transmit and
receive antenna pairs are assumed to be independent. We assume
independent noise at each receive antennas, samples of independent
circularly symmetric zero-mean complex Gaussian RVs with variance
$N_0/2$ per dimension.

In the remainder of this paper we will always assume $L_t=2$. It
will prove useful later to use the notation
\begin{equation}
\textbf{y}=\sqrt{\frac{E_{s}}{2}} \, \left[ \textbf{h}_{c1}\,
\textbf{h}_{c2} \right] \, \textbf{x}+\textbf{n} \label{mimomatr2}
\end{equation}
where $\textbf{h}_{ci}$ is the complex gain vector from the
$i^{th}$ transmit antenna to the receive antennas.

The present paper deals with a simplified yet optimal method to
estimate the transmit sequence \textbf{x}, i.e. it solves the ML
detection problem:
\begin{equation}
\tilde{\textbf{x}}=\arg \min_{\mathbf{x}}
\|\textbf{y}-\sqrt{\frac{E_s}{2}} \, \textbf{Hx}\|^2 \label{MLdem}
\end{equation}
The algorithms deals only with real quantities, i.e. the in-phase
(I) and quadrature-phase (Q) components of the complex quantities
in (\ref{mimomatr1}). To this end a suitable lattice
representation of the MIMO system is defined:
\begin{eqnarray}
\textbf{x}_r &=& \left[X_{1,I}, \ X_{1,Q}, \ X_{2,I}, \
X_{2,Q}\right]^T = \left[x_1,
\ldots , \, x_4\right]^T\\
\textbf{y}_r &=& \left[Y_{1,I}, \ Y_{1,Q}, \ldots \ , \,
Y_{L_r,I}, \ Y_{L_r,Q}\right]^T\\ \textbf{n}_r &=& \left[N_{1,I},
\ N_{1,Q}, \ldots \ , \,  N_{L_r,I}, \ N_{L_r,Q}\right]^T
\end{eqnarray}
Then (\ref{mimomatr2}) can be re-written as:
\begin{equation}
\textbf{y}_r=\sqrt{\frac{E_{s}}{2}} \, \textbf{H}_r \textbf{x}_r +
\textbf{n}_r = \sqrt{\frac{E_{s}}{2}} \, \left[\begin{array}{ccc}
\textbf{h}_1, \ldots , \, \textbf{h}_4 \end{array} \right] \,
\textbf{x}_r + \textbf{n}_r \label{lateqn}
\end{equation}
$\textbf{H}_r$ is the real channel matrix, which acts as the
lattice generator matrix - cfr. (\ref{sys_eq}). Each pair of
columns ($\textbf{h}_{2k-1}, \ \textbf{h}_{2k}$), $k=\{1,2\}$, has
the form:
\begin{eqnarray}
\textbf{h}_{2k-1} &=& \left[ \Re{\left[H_{1k}\right]}, \,
\Im{\left[H_{1k}\right]}, \ \Re{\left[H_{2k}\right]}, \,
\Im{\left[H_{2k}\right]}, \ldots , \, \Re{\left[H_{L_rk}\right]},
\, \Im{\left[H_{L_rk}\right]} \right]^T\\
\textbf{h}_{2k} &=& \left[ -\Im{\left[H_{1k}\right]}, \,
\Re{\left[H_{1k}\right]}, \ -\Im{\left[H_{2k}\right]}, \,
\Re{\left[H_{2k}\right]}, \ldots , \, -\Im{\left[H_{L_rk}\right]},
\, \Re{\left[H_{L_rk}\right]} \right]^T
 \label{hcol}
\end{eqnarray}
It should be noted that this ordering is slightly different than
the ordering used in the lattice search literature for multiple
antenna communications.  This change in ordering greatly impacts
the complexity and architecture of the ML demodulator.

\section{The Preprocessing Algorithm} \label{preproc}

This section describes an efficient way to preprocess
$\textbf{H}_r$, defined in (\ref{lateqn})-(\ref{hcol}). It should
be understood that a standard QR decomposition could be applied
without impairing the detection algorithm; the algorithm described
below however is more efficient as particular care is taken to
avoid performing unnecessary operations (e.g., vector
normalization, implying a real division and square root). Specifically for
$L_r\ge 2$ there is an $2L_t \times 2L_r$ orthogonal matrix
\begin{equation}
{\bf Q}=\left[\begin{array}{ccc} \textbf{h}_1 \quad \textbf{h}_2
\quad \textbf{q}_3 \quad \textbf{q}_4 \end{array} \right]
\end{equation}
where
\begin{eqnarray}
\textbf{q}_3 &=&
\|\textbf{h}_1\|^2\textbf{h}_3-(\textbf{h}_1^T\textbf{h}_3)\textbf{h}_1-
(\textbf{h}_2^T\textbf{h}_3)\textbf{h}_2\\
\textbf{q}_{4} &=&
\|\textbf{h}_1\|^2\textbf{h}_4-(\textbf{h}_1^T\textbf{h}_4)\textbf{h}_1-
(\textbf{h}_2^T\textbf{h}_4)\textbf{h}_2
\end{eqnarray}
such that
\begin{equation}
{\bf Q}^T{\bf Q}=\mbox{diag}\left[\left\|\textbf{h}_1\right\|^2,
\left\|\textbf{h}_1\right\|^2, \left\|\textbf{q}_3\right\|^2,
\left\|\textbf{q}_3\right\|^2 \right].
\end{equation}
It should be noted that
\begin{equation}
\left\|\textbf{q}_3\right\|^2=\left\|\textbf{h}_1\right\|^2\left(\left\|\textbf{h}_3\right\|^2
\left\|\textbf{h}_1\right\|^2
-\left(\textbf{h}_1^T\textbf{h}_3\right)^2-\left(\textbf{h}_2^T\textbf{h}_3\right)^2\right)=
\left\|\textbf{h}_1\right\|^2 r_3
\end{equation}
where
\begin{equation}
r_3=\left\|\textbf{h}_3\right\|^2 \left\|\textbf{h}_1\right\|^2
-\left(\textbf{h}_1^T\textbf{h}_3\right)^2-\left(\textbf{h}_2^T\textbf{h}_3\right)^2.
\end{equation}
There is a $2L_t \times 2L_t$ upper triangular matrix
\begin{equation}
{\bf R}=\left[\begin{array}{cccc}
1 & 0 & \textbf{h}_1^T\textbf{h}_3 & \textbf{h}_1^T\textbf{h}_4\\
0 & 1 & \textbf{h}_2^T\textbf{h}_3 & \textbf{h}_2^T\textbf{h}_4\\
0 & 0 & 1 & 0\\
0 & 0 & 0 & 1\end{array}\right]
\end{equation}
There is a $2L_t \times 2L_t$ diagonal matrix
\begin{equation}
{\bf \Lambda}_q=\mbox{diag}\left[1, 1, \
\left\|\textbf{h}_1\right\|^{-2}, \left\|\textbf{h}_1\right\|^{-2}
\right].
\end{equation}
These three matrices are related to the original real channel
matrix as
\begin{equation}
{\bf H}_r={\bf Q}{\bf R}{\bf \Lambda}_q.
\end{equation}
It should be noted that if $L_r=1$ then the bottom two rows of
${\bf R}$ will be eliminated but the same general form will hold
for the top two rows.

Because of this structure the detection problem on the MIMO
channel can be transformed into a structure suitable for lattice
search algorithms.  To this end note that the $4 \times 1$ vector
\begin{equation}
\tilde{\textbf{y}}={\bf Q}^T \textbf{y}_r=\sqrt{\frac{E_{s}}{2}}
\, {\bf \tilde{R}} \textbf{x}_r + {\bf Q}^T
\textbf{n}_r=\sqrt{\frac{E_{s}}{2}} \, \tilde{{\bf R}}
\textbf{x}_r + \tilde{\textbf{n}} \label{equivsys}
\end{equation}
where
\begin{equation} \label{eq:rtilde}
\tilde{{\bf R}}={\bf Q}^T{\bf Q}{\bf R}{\bf
\Lambda}_q=\left[\begin{array}{cccc}
\left\|\textbf{h}_1\right\|^2 & 0 & \textbf{h}_1^T\textbf{h}_3 & \textbf{h}_1^T\textbf{h}_4\\
0 & \left\|\textbf{h}_1\right\|^2 & \textbf{h}_2^T\textbf{h}_3 & \textbf{h}_2^T\textbf{h}_4\\
0 & 0 & r_3 & 0\\
0 & 0 & 0 & r_3\end{array}\right].
\end{equation}
The noise vector in the triangular model still has independent
components but the components have unequal variances, i.e.,
\begin{equation}
{\bf
R}_{\tilde{n}}=E\left[\tilde{\textbf{n}}\tilde{\textbf{n}}^T\right]
= \frac{N_0}{2}\,\mbox{diag}\left[\left\|\textbf{h}_1\right\|^2, \
\left\|\textbf{h}_1\right\|^2, \ \left\|\textbf{h}_1\right\|^2
r_3, \ \left\|\textbf{h}_1\right\|^2 r_3 \right].
\end{equation}
The interesting characteristic of the model formulation in this
manner is that each of the I and Q components of each transmitted
signal are broken into orthogonal dimensions and can be searched
in an independent fashion.

All parameters needed in this triangularized model are a function
of eight variables. Four of the variables are functions of the
channel only, i.e.,
\begin{equation}
\sigma^2_1=\left\|\textbf{h}_1\right\|^2 \qquad
\sigma^2_2=\left\|\textbf{h}_3\right\|^2 \qquad
s_1=\textbf{h}_1^T\textbf{h}_3 \qquad
s_2=\textbf{h}_1^T\textbf{h}_4.
\end{equation}
and four are functions of the channel and the observations, i.e.,
\begin{equation}
V_1=\textbf{h}_1^T\textbf{y}_r \qquad
V_2=\textbf{h}_2^T\textbf{y}_r \qquad
V_3=\textbf{h}_3^T\textbf{y}_r \qquad
V_4=\textbf{h}_4^T\textbf{y}_r.
\end{equation}
It should be noted that
$\left\|\textbf{h}_1\right\|^2=\left\|\textbf{h}_2\right\|^2$,
$\left\|\textbf{h}_3\right\|^2=\left\|\textbf{h}_4\right\|^2$,
$\textbf{h}_1^T\textbf{h}_3=\textbf{h}_2^T\textbf{h}_4$, and that
$\textbf{h}_1^T\textbf{h}_4=-\textbf{h}_2^T\textbf{h}_3$.  These
two equalities imply that the $2 \times 2$ matrix in the upper
right corner of ${\bf \tilde{R}}$ is a rotation matrix.
Specifically the required results for the upper triangular
formulation is
\begin{equation} \label{eq:utriangular}
\tilde{\textbf{y}}=\left[\begin{array}{c}
\tilde{y_1} \\
\tilde{y_2} \\
\tilde{y_3} \\
\tilde{y_4}\end{array} \right] = \left[\begin{array}{c}
V_1 \\
V_2 \\
\sigma^2_1V_3-s_1V_1+s_2V_2 \\
\sigma^2_1V_4-s_2V_1-s_1V_2\end{array}\right] \quad {\bf
\tilde{R}}=\left[\begin{array}{cccc}
\sigma^2_1 & 0 & s_1 & s_2\\
0 & \sigma^2_1 & -s_2 & s_1\\
0 & 0 & \sigma^2_1\sigma^2_2-s_1^2-s_2^2 & 0 \\
0 & 0 & 0 & \sigma^2_1\sigma^2_2-s_1^2-s_2^2
\end{array}\right].
\end{equation}
This formulation greatly simplifies the lattice search formulation
and results in a preprocessing complexity that is $O(16L_r)$.

As it will prove useful when dealing with soft output generation,
shifting the ordering of the transmit antennas will result in a similar
model. When the order of transmit antennas is reversed the model becomes

\begin{equation}
\tilde{\textbf{y}}_s=\left[\begin{array}{c}
\tilde{y_{s1}} \\
\tilde{y_{s2}} \\
\tilde{y_{s3}} \\
\tilde{y_{s4}}\end{array} \right]=\left[\begin{array}{c}
V_3 \\
V_4 \\
\sigma^2_2V_1-s_1V_3-s_2V_4 \\
\sigma^2_2V_2+s_2V_3-s_1V_4
\end{array}\right]
\, {\bf \tilde{R}}_s=\left[\begin{array}{cccc}
\sigma^2_2 & 0 & s_1 & -s_2\\
0 & \sigma^2_2 & s_2 & s_1\\
0 & 0 & \sigma^2_1\sigma^2_2-s_1^2-s_2^2 & 0 \\
0 & 0 & 0 & \sigma^2_1\sigma^2_2-s_1^2-s_2^2
\end{array}\right]
\label{shifteq}
\end{equation}

\section{ML Demodulation} \label{ML_demod}

In this section we describe how the system equations defined in
Section~\ref{ML_demod} lead to a simplified yet optimal ML
demodulation. Consider a PSK or QAM constellation of size {\it S}.
For the sake of conciseness, the discussion here will assume that
$(M^2)$-QAM modulation is used on each antenna, but the derivation
is valid - with straightforward generalizations - for any complex
constellation. The optimum ML word demodulator (\ref{MLdem}) would
have to compute the ML metric for $M^{2L_t}$ constellation points
and has a complexity $O(M^4)$ for $L_t=2$.

The notation used in the sequel is that $\Omega_x$ will refer to
the {\it M}-PAM constellation for each real dimension.  Given the
formulation in (\ref{equivsys})-(\ref{eq:utriangular}) and
neglecting scalar energy normalization factors for simplicity, the
ML decision metric becomes
\begin{eqnarray}
T(\textbf{x}_r)&=&-\frac{\left(\tilde{y}_1-\sigma^2_1x_{1}-s_1x_{3}-s_2x_{4}\right)^2}{\sigma^2_1}
-\frac{\left(\tilde{y}_2-\sigma^2_1x_{2}+s_2x_{3}-s_1x_{4}\right)^2}{\sigma^2_1}
\nonumber \\ &&
-\frac{\left(\tilde{y}_3-r_3x_{3}\right)^2}{\sigma^2_1r_3}
-\frac{\left(\tilde{y}_4-r_3x_{4}\right)^2}{\sigma^2_1r_3}
\label{LORDmetric}
\end{eqnarray}
The ML demodulator finds the maximum value of the metric over all
possible values of the sequence $\textbf{x}_r$.  This search can
be greatly simplified by noting for given values of $x_{3}$ and
$x_{4}$ the maximum likelihood metric reduces to
\begin{equation} \label{eq:condmlwd}
T(\textbf{x}_r)=-\frac{\left(\tilde{y}_1-\sigma^2_1x_{1}-C_1(x_3,x_4)\right)^2}{\sigma^2_1}
-\frac{\left(\tilde{y}_2-\sigma^2_1x_{2}-C_2(x_3,x_4)\right)^2}{\sigma^2_1}
- C_3(x_3,x_4) \label{LORDmetric2}
\end{equation}
where
\begin{equation}
C_1(x_3,x_4)=s_1x_3+s_2x_4 \qquad C_2(x_3,x_4)=-s_2x_3+s_1x_4
\qquad C_3(x_3,x_4) \ge 0
\end{equation}
It is clear from examining (\ref{eq:condmlwd}) that due to the
orthogonality of the problem formulation the conditional ML
decision on $x_1$ and $x_2$ can immediately be made by a simple
threshold test, i.e.,
\begin{equation} \label{eq:slice}
\hat{x}_1(x_3,x_4)=\mbox{round}\left(\frac{\tilde{y}_1-C_1(x_3,x_4)}{\sigma^2_1}\right),
\quad
\hat{x}_2(x_3,x_4)=\mbox{round}\left(\frac{\tilde{y}_2-C_2(x_3,x_4)}{\sigma^2_1}\right).
\end{equation}
The round operation is a simple slicing operation to the
constellation elements of $\Omega_x$.  The final ML  estimate is
then given as
\begin{eqnarray}\label{eq:mlwd1}
&&
\left\{\hat{x_1}(\hat{x_3},\hat{x_4}),\hat{x_2}(\hat{x_3},\hat{x_4}),\hat{x_3},\hat{x_4}\right\}=
\arg \stackrel{\textstyle \max}{\scriptstyle x_3,x_4 \in
\Omega_x^2}\left\{-\frac{\left(\tilde{y}_1-\sigma^2_1\hat{x}_1(x_3,x_4)-C_1(x_3,x_4)\right)^2}{\sigma^2_1}
\nonumber\right. \\ &&
-\left.\frac{\left(\tilde{y}_2-\sigma^2_1\hat{x}_2(x_3,x_4)-C_2(x_3,x_4)\right)^2}{\sigma^2_1}
 - C_3(x_3,x_4) \right\}
\end{eqnarray}
This implies that the number of points that has to be searched in
this formulation to find the true ML estimator is $M^2$ (with two
slicing operations per searched point) and not $M^4$. This is a
significant saving in complexity.

Examining (\ref{eq:slice}) and (\ref{eq:mlwd1}) shows this reduced
complexity ML demodulation is a direct consequence of the
reordered lattice formulation. Recall each group of two rows in
the model correspond to a transmit antenna. Equation
(\ref{eq:slice}) shows that at the top of the triangularized model
the decisions for the first transmit antenna can be made
independently for the I and the Q modulation.  This was not true
for the traditional lattice formulation \cite{Damen} as after the
triangularization the higher rows become dependent on all the
lower layers of the transmit modulation.  Retaining this
orthogonalization greatly simplifies the optimal search and has
important implications for suboptimal searches.  Secondly this
orthogonalization also greatly facilitates parallel searches,
solving one of SD drawbacks, i.e. the fact that the search must be
performed in a recursive fashion.

\section{LLR generation} \label{LLR}

The problem is first here recalled for complex-domain system
(\ref{mimomatr1}). Consider the information symbol vector
\textbf{x} $=(X_1 \, \, X_2)^{T}$, where $X_j,\,j=1,\,2$ is a
complex symbol belonging to a given $M^2$-QAM constellation and be
$M_c$ the number of bits per symbol. The bit soft-output
information is the a-posteriori probability (APP) ratio of the bit
$b_k$, $k=1, \ldots , 2M_c\,$, conditioned on the received channel
symbol vector \textbf{y}; that is often expressed in the
logarithmic domain (log-likelihood ratio, LLR) as:
\begin{equation}
L(b_k|\textbf{y})=\ln{\frac{P(b_k=1|\textbf{y})}{P(b_k=0|\textbf{y})}}=\ln{\frac{\displaystyle\sum_{{\bf
x} \in
S(k)^+}P(\textbf{y}|\textbf{x})P_a(\textbf{x})}{\displaystyle\sum_{{\bf
x} \in S(k)^-}P(\textbf{y}|\textbf{x})P_a(\textbf{x})}}
\label{generalLLR}
\end{equation}
where $S(k)^+$ is the set of $2^{2Mc-1}$ bit sequences having
$b_k=1$, and similarly $S(k)^-$ is the set of bit sequences having
$b_k=0$; $P_a({\bf x})$ represent the a-priori probabilities of
{\bf x}, which can be neglected in case of equiprobable transmit
symbols, as it is the case of this paper. In the general case, the
likelihood function $\displaystyle P(\textbf{y}|\textbf{x})$ can
be derived from (\ref{mimomatr1}):
\begin{equation}
P(\textbf{y}|\textbf{x})\propto
\exp{\left[-\frac{1}{2\sigma^2}\|\textbf{y}-\sqrt{\frac{E_s}{2}}\textbf{Hx}\|^2
\right]}=\exp{\left[-D(\textbf{x})\right]}
\end{equation}
where $\sigma^2=N_0/2$ and $D(\textbf{x})$ is the Euclidean
distance term. The summation of exponentials involved in
(\ref{generalLLR}) is often approximated according to the
following so-called max-log approximation:
\begin{equation}
\ln{\sum_{{\bf x}\in S(k)^+} \exp{\left[-D({\bf
x})\right]}}\approx\ln{\max_{{\bf x}\in S(k)^+}{\exp{\left[-D({\bf
x})\right]}}}=-\min_{{\bf x}\in S(k)^+}{D({\bf x})} \label{max}
\end{equation}
Alternatively, it is possible to exactly compute
(\ref{generalLLR}) through the ``Jacobian logarithm'' or $\max^*$
function
\begin{equation}
\textrm{jacln}(a,b):=\ln{\left[\exp{(a)}+\exp{(b)}\right]}=\max{(a,b)}+\ln{\left[1+\exp{(-|a-b|)}\right]}.
\label{maxstar}
\end{equation}
Simulations in \cite{Robertson} show that the performance
degradation due to max-log approximation is generally very small
compared to the use of $\max^*$ function. Using (\ref{max}),
(\ref{generalLLR}) can then be written as:
\begin{equation}
L(b_k|\textbf{y})\approx\min_{{\bf x}\in
S(k)^-}{D(\textbf{x})}-\min_{{\bf x}\in S(k)^+}{D(\textbf{x})}
\label{maxlogLLR}
\end{equation}
In the sequel we will show how LORD can provide an exact
computation of (\ref{maxlogLLR}) but with a much lower complexity
than exhaustive search-ML. The computation of (\ref{maxlogLLR})
requires identification of the most likely lattice point with
$b_k=1$ and the most likely lattice point with $b_k=0$ for each
bit index $k=1,\ldots,2M_c$. The problem in the case of the SD
algorithm is clearly stated in \cite{ten_Brink}. By definition,
one of the two sequences is the (optimum) hard-decision ML
solution of (\ref{MLdem}). However, using SD, there is no
guarantee that the other sequence is one of the valid lattice
points found by SD during the process of the lattice search.

LORD does not have this problem generating LLRs. To show this let
us consider the bits corresponding to the complex symbol $X_2$ in
the symbol sequence \textbf{x} $=(X_1 \, \, X_2)^{T}$. After the
lattice representation is derived, from (\ref{equivsys}) and
(\ref{LORDmetric}) the likelihood function $\displaystyle
P(\tilde{\textbf{y}}|\textbf{x}_r)$ is given by:
\begin{equation}\label{lklfctn}
P(\tilde{\textbf{y}}|\textbf{x}_r)=\exp\left[-|T(\textbf{x}_r)|\right].
\end{equation}
where $T(\textbf{x}_r)$ is defined in (\ref{LORDmetric}). Using
arguments similar to those that led to the simplified ML
demodulation (\ref{LORDmetric2})-(\ref{eq:slice}), one can easily
prove that the two sequences needed for every bit in $X_2$ are
certainly found minimizing (\ref{LORDmetric}) over the possible
$M^2$ values of ($x_3, \, x_4$) and performing a simple slicing
operation to the constellation elements of $\Omega_x$; thus the
desired couples ($x_1, \, x_2$) are uniquely determined for every
($x_3, \, x_4$). Equation (\ref{generalLLR}) can be then written
as:
\begin{equation}
L(b_{2,k}|\tilde{\textbf{y}})\approx\min_{x_3,x_4\in
S(k)_2^-}{T(\textbf{x}_r)}-\min_{x_3,x_4\in
S(k)_2^+}{T(\textbf{x}_r)} \label{LLR_X2}
\end{equation}
where $b_{2,k}$ are the bits corresponding to $X_2$, $k=1, \ldots
, M_c$, and $S(k)_2^+$ ($S(k)_2^-$) are the set of $2^{M_c-1}$ bit
sequences having $b_{2,k}=1$ ($b_{2,k}=0$).

The computation of the LLRs for the bits corresponding to symbols
in ${x_1, x_2}$ can be obtain by a simple reordering of the model
and a repeating of the LORD processing.   To this end denote
$\textbf{x}_s=\left[x_3 \, x_4 \, x_1 \, x_2 \right]$ as the
reordered real modulation symbols, then the LLR can be given as
\begin{equation}
L(b_{1,k}|\tilde{\textbf{y}_s})\approx\min_{x_1,x_2\in
S(k)_1^-}{T'(\textbf{x}_s)}-\min_{x_1,x_2\in
S(k)_1^+}{T'(\textbf{x}_s)} \label{LLR_X1}
\end{equation}
where $b_{1,k}$ are the bits corresponding to $X_1$, $k=1, \ldots
, M_c$, $S(k)_1^+$ ($S(k)_1^-$) are the set of $2^{M_c-1}$ bit
sequences having $b_{1,k}=1$ ($b_{1,k}=0$). The reordered ML
decision metric becomes
\begin{eqnarray}
T'(\textbf{x}_s)&=&-\frac{\left(\tilde{y}_{s1}-\sigma^2_2x_{3}-s_1x_{1}+s_2x_{2}\right)^2}{\sigma^2_2}
-\frac{\left(\tilde{y}_{s2}-\sigma^2_2x_{4}-s_2x_{1}-s_1x_{2}\right)^2}{\sigma^2_2}
\nonumber \\ &&
-\frac{\left(\tilde{y}_{s3}-r_3x_{1}\right)^2}{\sigma^2_2r_3}
-\frac{\left(\tilde{y}_{s4}-r_3x_{2}\right)^2}{\sigma^2_2r_3}
\label{shiftedLORDmetric}
\end{eqnarray}
in a direct analogy to (\ref{shifteq}). By comparing
(\ref{shifteq}) and (\ref{eq:utriangular}), it should be clear
that computing the processing required by the reordered sequence
involves a low amount of extra-complexity. Besides, the LLR
computation for the bits corresponding to symbols $X_1$ and $X_2$
can be carried out in a parallel fashion. Thus, we have derived an
exact Max-log bit LLR computation using two layer orderings and an
overall lattice search over $2M^2$ sequences instead of $M^4$ as
would be required by the exhaustive search-based ML algorithm.

\section{Performance results} \label{perf}
\begin{figure}
\begin{minipage}[t]{3.1in}
\centering{\includegraphics[width=3.1in]{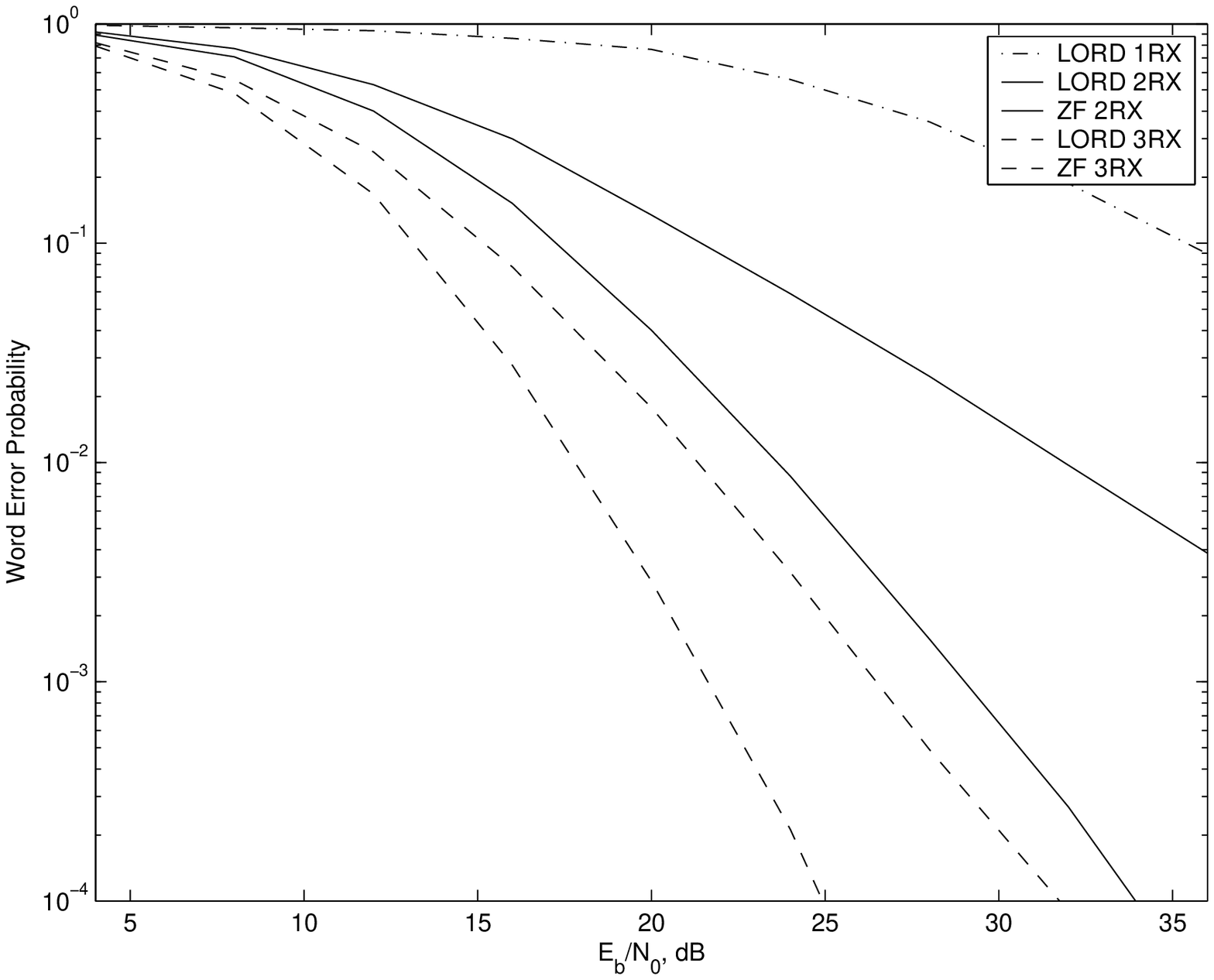}} a) Word error
probability.
\end{minipage}
\begin{minipage}[t]{3.1in}
\centering{\includegraphics[width=3.1in]{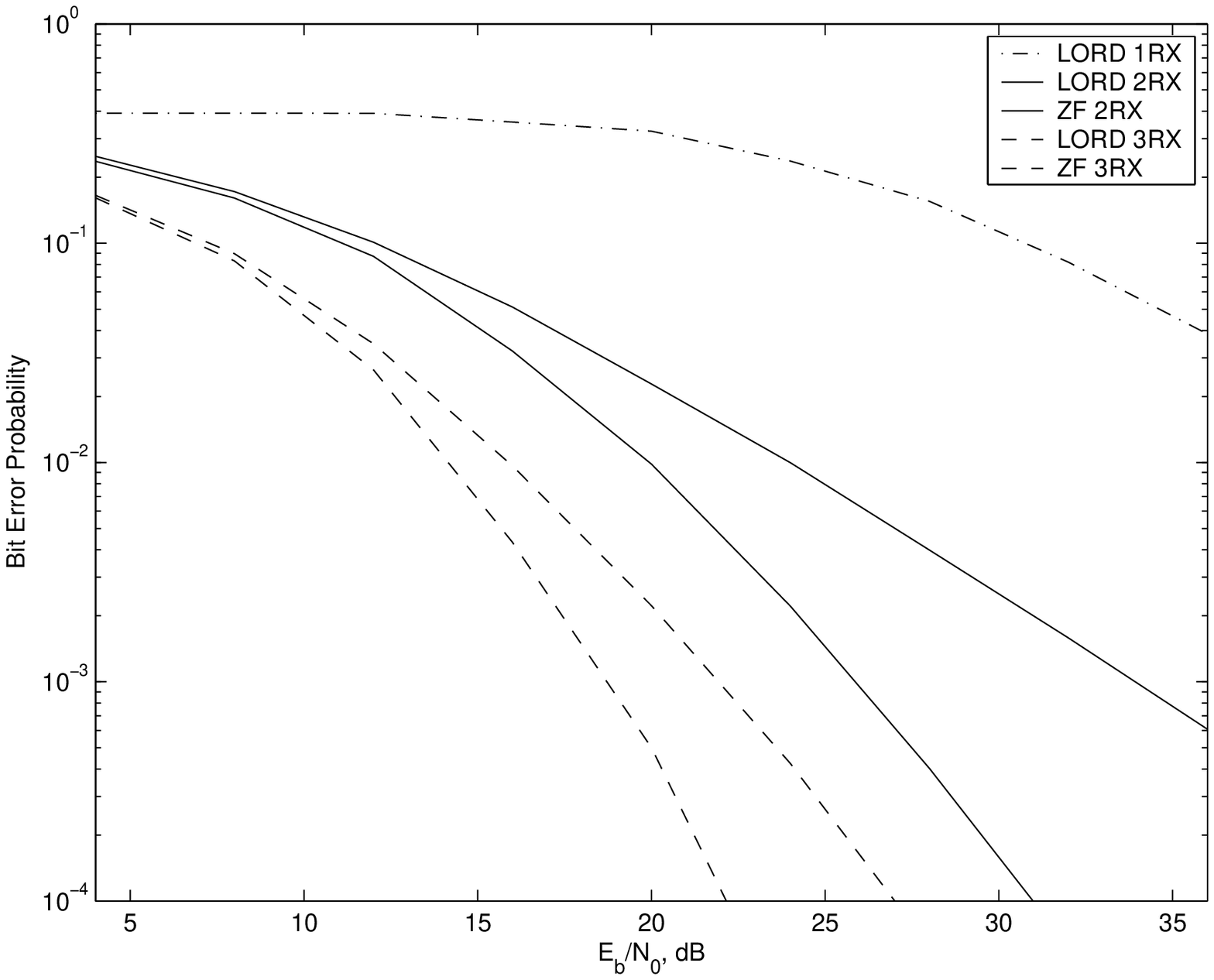}} b) Bit error
probability.
\end{minipage}
\caption{The decoding performance for uncoded MIMO transmission. $L_t=2$
and 12 bits per channel use.}
\label{fig:MLperf}
\end{figure}

In this section two examples are presented with the corresponding
performance evaluated by simulation.  The examples will be limited to
frequency flat and time flat fading for simplicity but this does not
represent a limit on how LORD can be applied in MIMO detection.  The first
example is an uncoded
$L_t=2$ system using 64QAM modulation.  The maximum likelihood detection
word error probability performance in spatially white Rayleigh fading is
shown in Fig.~\ref{fig:MLperf}.  The ML performance is compared to the
well understood zero forcing (ZF) detector. The advantage of ML detection
versus linear detection in terms of diversity is obvious from these plots.

\begin{figure}
\begin{minipage}[t]{3.1in}
\centering{\includegraphics[width=3.1in]{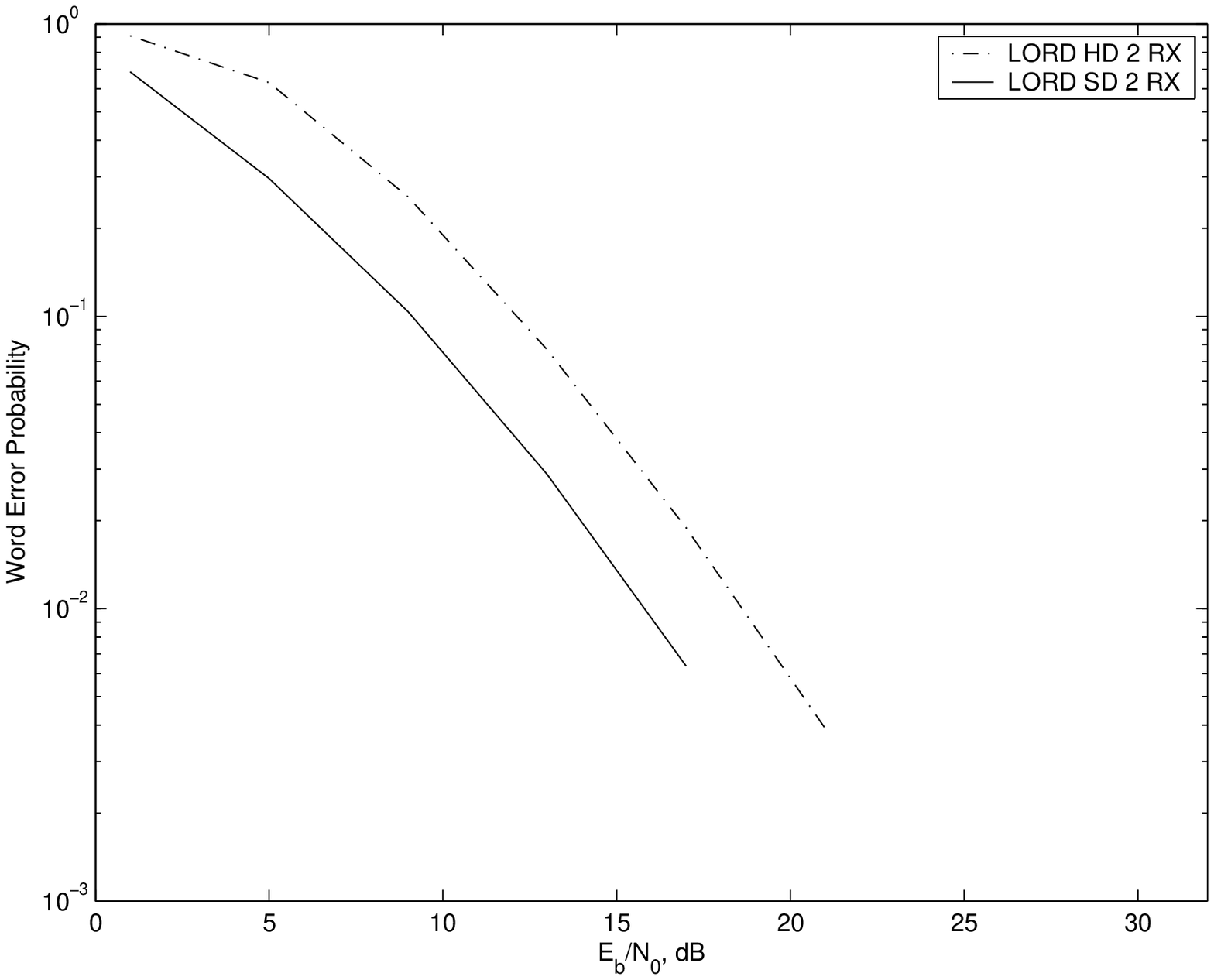}} a) Word error
probability.
\end{minipage}
\begin{minipage}[t]{3.1in}
\centering{\includegraphics[width=3.1in]{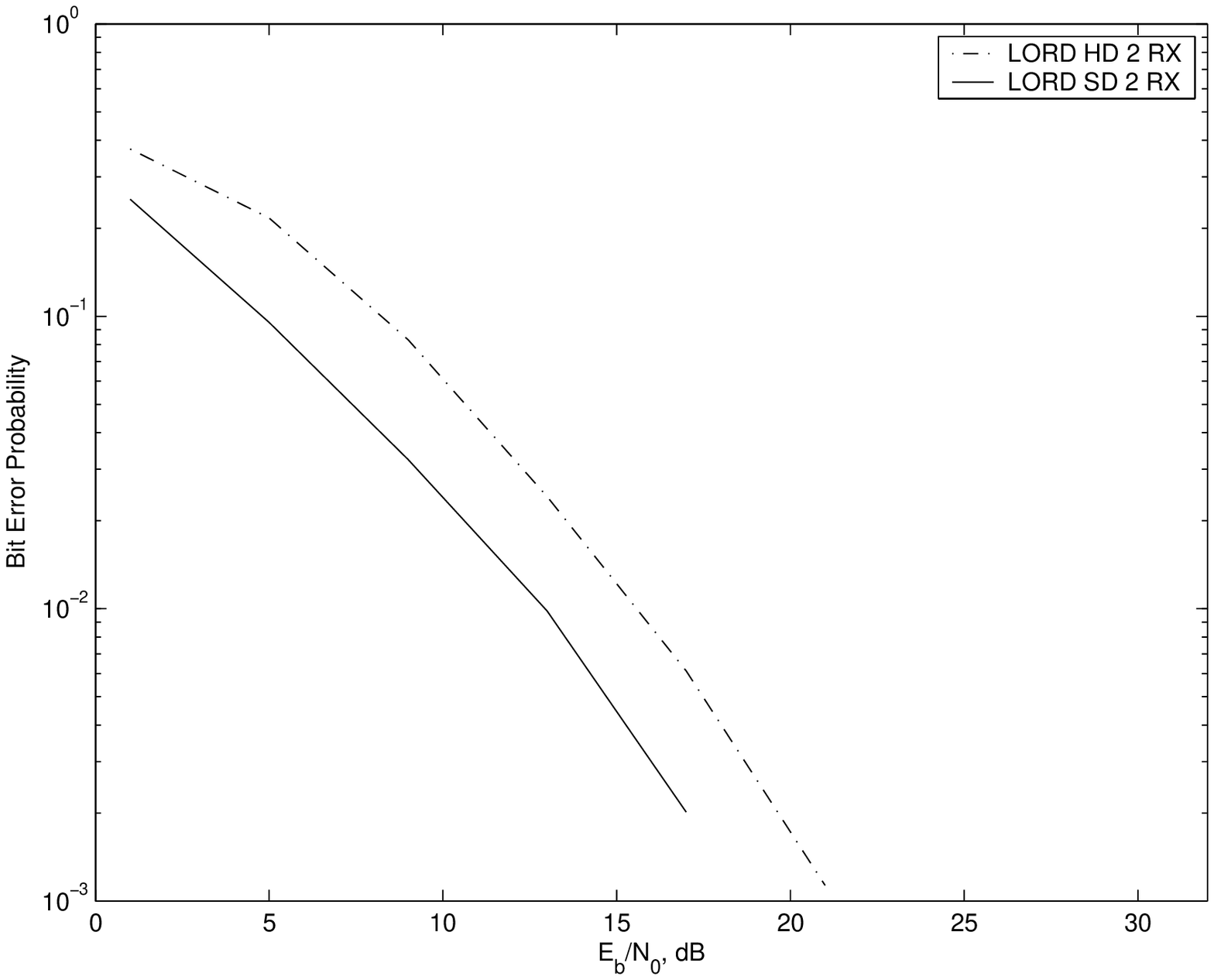}} b) Bit error
probability.
\end{minipage}
\caption{The BICM decoding performance. $L_t=2$, $L_R=2$ and 6
bits per channel use.}
\label{fig:BICMperf}
\end{figure}

A second system consists of a bit interleaved coded modulation
(BICM) with a frame size of 144 coded bits using 64QAM modulation
on each antenna.  The interleaver used was a $12 \times 12$ block
interleaver so that each adjacent bit could be permuted to a
different antenna and different significant bit on the QAM
modulation while being on different time slots.  The convolutional
code is the standard 64-state rate 1/2 binary convolutional code
with octal generators (133, 171). The performance is shown in
Fig.~\ref{fig:BICMperf} for $L_t=2$, $L_r=2$, and spatially white
Rayleigh fading for both hard and max-log inner bit metric
generation.  Clearly this figure demonstrates LORD is a system
that can exactly compute the max-log LLR at a low complexity when
used in a concatenated MIMO coded modulation.

\section{Conclusion} \label{concl}
We have presented a novel lattice search-based MIMO detection
algorithm for two transmit antennas, characterized by a low
preprocessing complexity, that achieves optimal ML demodulation
with a complexity of the order of $S$, if $S$ is the constellation
size. Also, the algorithm is able to generate optimal Max-log bit
LLRs using a parallel lattice search over $2S$ transmit sequences.

To our knowledge no ML-approaching detection technique, among
those proposed so far, is able to generate a low complexity optimal bit
soft output information and easily be suitable for a parallel hardware
implementation. LORD provides all these desirable features and thus
represents a significant improvement over the state of the art in this
field.

\end{document}